\documentclass[12pt]{amsart}
\usepackage{amscd}
\usepackage{amsfonts}
\usepackage{epsf}
\usepackage{epic}
\usepackage{eepic}
\usepackage{latexsym}

\begin{document}

\title[Localization of Equivariant Cohomology]
{Localization of Equivariant Cohomology - Introductory and Expository
Remarks}  
\author{A.A. Bytsenko}
\address{Departamento de Fisica, Universidade Estadual de Londrina,
Caixa Postal 6001, Londrina-Parana, Brazil\,\, {\em E-mail address:} 
{\rm abyts@fisica.uel.br}}
\author{F.L. Williams}
\address{Department of Mathematics, University of Massachusetts,
Amherst, Massachusetts 01003, USA\,\, {\em E-mail address:} 
{\rm williams@math.umass.edu}}

\date{August, 2000}

\thanks{First author partially supported by a CNPq grant (Brazil), RFFI 
grant (Russia) No 98-02-18380-a, and by GRACENAS grant (Russia) No 6-18-1997.}

\maketitle

%\begin{abstract}
%\end{abstract}

\section{Introduction}

In 1982 J.J. Duistermaat and G. Heckman \cite{duistermaat} found a 
formula which expressed
certain oscillatory integrals over a compact symplectic manifold as a sum over
critical points of a corresponding phase function. In this sense these
integrals are localized, and their stationary-phase approximation is exact 
with no error terms occuring. The ideas and techniques of localization
extended to infinite-dimensional settings have proved to be quite useful and
indeed central for many investigations in theoretical physics - investigations
ranging from supersymmetric quantum mechanics, topological and supersymmetric
field theories, to integrable models and low-dimensional gauge theories,
including two-dimensional Yang-Mills theory \cite{szabo}. Path integral 
localization appears in the work of M. Semenov-Tjan-Schanskii 
\cite{semenov}, which actually pre-dates \cite{duistermaat}.

E. Witten was the first to propose an extension of the Duistermaat-Heckman
(D-H) formula to an infinite-dimensional manifold - namely to the loop space
$LM$ of smooth maps from the circle ${\Bbb S}^1$ to a compact orientable 
manifold $M$. In this case a purely formal application of the D-H formula to 
the partition function of $N=1/2$ supersymmetric quantum mechanics yields a 
correct formula for the index of a Dirac operator \cite{atiyah1}. 
Further arguments in this direction were presented with mathematical rigor 
by J.-M. Bismut in \cite{bismut1,bismut2}.

The various generalizations of D-H generally require formulations in terms of 
equivariant cohomology. One has, for example, the Berline-Vergne (B-V)
localization formula \cite{berline1,berline2,berline3,berline4} which 
expresses the integral of an equivariant 
cohomology class as a sum over zeros of a vector field to which that class is 
related; also see \cite{blau,niemi1,szabo,woodhouse} for example, a broder 
formulation of the localization formula.
Our remarks here are designed to provide members of the Conference, and others,
with a brief introduction to the B-V localization formula, and to indicate how
the D-H formula is derived from it. Thus our goal is deliberately very
modest. We shall limit our discussion, in particular, to the 
finite-dimensional setting as our idea is to convey the basic flavor of 
these formulas. This introduction should prepare readers for quite more 
ambitions discussions found in \cite{berline4,niemi1,szabo}, for example.

The role of equivariant cohomology in physical theories will continue to
grow as it has grown in past years. In particular it will be an indispensable 
tool for topological theories of gauge, strings, and gravity. We thank the 
organizers of this Conference for this opportunity to present these brief
remarks on a topic of such growing interest in the physics community.

\section{The equivariant cohomology space $H(M,X,s)$}

For an integer $j\geq 0$ let $\Lambda^jM$ denote the space of smooth complex
differential forms of degree $j$ on a smooth manifold $M$.
$d:$ $\Lambda^jM\rightarrow \Lambda^{j+1}M$ will denote exterior 
differentiation, and for a smooth vector field $X$ on $M$, 
$\theta(X): \Lambda^jM\rightarrow\Lambda^jM,\,\, i(X): \Lambda^jM\rightarrow 
\Lambda^{j-1}M$ will denote Lie and interior differentiation by $X$,
respectively:

$$
(\theta(X)\omega)(X_1,..., X_j)=X\omega(X_1,..., X_j)
$$
$$
-\sum_{\ell=1}^j\omega(X_1,..., X_{\ell-1}, [X,X_\ell], X_{\ell+1},..., X_j)
\mbox{,}
\eqno{(2.1)}
$$

$$
(i(X)\omega)(X_1,..., X_{j-1})=\omega(X,X_1,... X_{j-1})
\eqno{(2.2)}
$$
for $\omega\in \Lambda^jM$ and for $X_1,..., X_j \in VM$ = the space of
smooth vector fields on $M$. One has the familiar rules

$$
\theta(X)=di(X)+i(X)d
\mbox{,}
$$
$$
d\theta(X)=\theta(X)d, \,\,\, \theta(X)i(X)=i(X)\theta(X)
\mbox{,}
$$
$$
i(X)^2=0; \,\,\,{\rm of\,\,\, course}\,\,d^2=0
\mbox{.}
\eqno{(2.3)}
$$
For a complex number $s$ let

$$
d_{X,s}=d+si(X)\,\,\,{\rm on}\,\,\Lambda M=\sum_{j\geq0}\oplus\Lambda^jM
\mbox{.}
\eqno{(2.4)}
$$
Then by (2.3), $d_{X,s}\theta(X)=\theta(X)d_{X,s}$ and $d_{X,s}^2=s\theta(X)$.
Hence the subspace

$$
\Lambda_XM=\{\omega \in \Lambda M | \theta(X)\omega=0\}
\mbox{,}
\eqno{(2.5)}
$$
of $\Lambda M$ is $d_{X,s}-$ invariant and $d_{X,s}^2=0$ on $\Lambda_XM$.
It follows that we can define the cohomology space

$$
H(M,X,s)=Z(M,X,s)/B(M,X,s)
\eqno{(2.6)}
$$
for $Z(M,X,s)=$ kernel of $d_{X,s}$ on $\Lambda_XM,\,\, \, 
B(M,X,s)=d_{X,s}\Lambda_X M$. The space $H(M,X,s)$ appears to depend on the 
parameter $s$. However it is not difficult to show that for $s\neq 0$ there is
an isomorphism of $H(M,X,s)$ onto $H(M,X,1)$. For $X=0$, $H(M,0,s)$ is the
ordinary de Rham cohomology of $M$.

We shall be interested in the case when M has a smooth Riemannian structure 
$< , >$, and when $M$ is oriented and even-dimensional. Thus let 
$\omega\in \Lambda^{2n}M-\{0\}$, ${\rm dim}\,M=2n$, define the orientation of
$M$. In this case we assume moreover that $X$ is a Killing vector field:

$$
X<X_1, X_2>\, = \,<[X, X_1], X_2>+<X_1, [X, X_2]>
\eqno{(2.7)}
$$
for $X_1, X_2 \in VM$. If $p\in M$ is a zero of $X$ (i.e. $X_p=0$) then there
is an induced linear map ${\frak L}_p(X)$ of the tangent space $T_p(M)$ of
$M$ at $p$ such that

$$
{\frak L}_p(X)(Z_p)=[X, Z]_p\,\,\,\,\,{\rm for}\,\,\,Z\in VM
\mbox{.}
\eqno{(2.8)}
$$
Because of\,\, (2.7)\,\, one has that\,\,\, ${\frak L}_p(X)$\,\,\, is\,\,\, 
skew-symmetric;\,\,\, i.e.\,\,\, 
$<{\frak L}_p(X)V_1, V_2>_p\, =-<V_1,{\frak L}_p(X)V_2>_p$\,\, for 
$V_1,V_2 \in T_p(M)$. Let $f_p(X): T_p(M)\oplus T_p(M)\rightarrow {\Bbb R}$
be the corresponding skew-symmetric bilinear form on $T_pM$:

$$
f_p(X)(V_1,V_2) =\,<V_1,{\frak L}_p(X)V_2>_p\,\,\,{\rm for}\,\,\,V_1,V_2\in
T_pM
\mbox{.}
\eqno{(2.9)}
$$

In order to apply some standard linear algebra to the real inner product 
space $(T_p(M),\, < , >_p)$, we suppose ${\frak L}_p(X)$ is a non-singular 
linear
operator on $T_p(M): {\rm det}{\frak L}_p(X)\neq 0$; equivalently, this means
that the bilinear form $f_p(X)$ is non-degenerate. Then one can find an ordered
orthonormal basis $e=e^{(p)}=\{e_j=e_j^{(p)}\}_{j=1}^{2n}$ of $T_p(M)$ such
that

$$
\!\!\!\!\!\!\!\!\!\!\!\!\!\!\!\!\!\!\!\!\!\!\!\!\!
\!\!\!\!\!\!\!\!\!\!\!\!\!\!\!\!\!\!\!
{\frak L}_p(X)e_{2j-1}=\lambda_je_{2j}
\mbox{,}
$$
$$
{\frak L}_p(X)e_{2j}=-\lambda_je_{2j-1},\,\,\,\,\,\,{\rm for}\,\,\,
1\leq j\leq n
\mbox{,}
\eqno{(2.10)}
$$
where each $\lambda_j\in {\Bbb R}-\{0\}$. In other words, relative to $e$
the matrix of ${\frak L}_p(X)$ has the form

$$
{\frak L}_p(X) = \left[ \begin{array}{ccccccc}
0 & -\lambda_1 &  &  &  &  &    \\
\lambda_1 & 0  &  &  &  &  &    \\
            &  &  . &  &  &  &    \\
            &  &  & . &  &  &    \\
           &  &  &  & . &  &    \\
  &  &  &  &  &  0 & -\lambda_n \\
  &  &  &  &  &  \lambda_n & 0  \\ 
\end{array} \right]
\mbox{.}
\eqno{(2.11)}
$$    
Moreover, interchanging $e_1, e_2$ if necessary, we can assume that $e$
is positively oriented: $\omega_p(e_1,..., e_{2n})>0$. Finally, consider
the Pfaffian ${\rm Pf}_e({\frak L}_p(X))$ of ${\frak L}_p(X)$ relative
to $e$:

$$
{\rm Pf}_e({\frak L}_p(X))=\frac{1}{n!}\left[f_p(X)\wedge...\wedge
f_p(X)\right](e_1,...,e_{2n})
\mbox{.}
\eqno{(2.12)}
$$
${\rm Pf}_e({\frak L}_p(X))$ satisfies 

$$
\!\!\!\!\!\!\!\!\!\!
({\rm *})\,\,\,\,\,\,\,\, {\rm Pf}_e({\frak L}_p(X))^2
={\rm det}{\frak L}_p(X),
$$
$$
({\rm **})\,\,\,\,\,\,\,\, {\rm Pf}_e({\frak L}_p(X))= 
(-1)^n\lambda_1\cdot\cdot\cdot \lambda_n.
$$
If $e'=\{e_{j}'\}_{j=1}^{2n}$ is another ordered, positively oriented
orthogonal basis of $T_p(M)$ then

$$
{\rm Pf}_{e'}({\frak L}_p(X))={\rm Pf}_e({\frak L}_p(X))
\mbox{.}
\eqno{(2.13)}
$$
Equation (2.13) means that we can define a square-root of ${\frak L}_p(X)$
by setting

$$
\left[{\rm det}{\frak L}_p(X)\right]^{1/2}=(-1)^n{\rm Pf}_e({\frak L}_p(X))
\mbox{.}
\eqno{(2.14)}
$$
That is, the square-root is independent of the choice $e$ of an ordered, 
positively oriented orthogonal basis of $T_p(M)$. 
By\,\,\, $({\rm **})$\,\,\, we have\,\,\,
$[{\rm det}{\frak L}_p(X)]^{1/2}=\lambda_1\cdot\cdot\cdot \lambda_n$. The
reader is reminded that the hypotheses $X_p=0$ and 
${\rm det}({\frak L}_p(X))\neq 0$ were imposed, with $X$ a Killing vector 
field.

\section{The localization formula}

As before we are given an oriented, $2n-$ dimensional Riemannian manifold
$(M,\omega,< , >)$. Now assume that $G$ is a compact Lie group which acts
smoothly on $M$, say on the left, and that the metric $< , >$ is $G-$
invariant. Let ${\rm {\bf g}}$ denote the Lie algebra of $G$. Given 
$X\in {\rm {\bf g}}$, there
is an induced vector field $X^{*}\in VM$ on $M$: for $\phi\in C^{\infty}(M)$,
$p\in M$

$$
(X^{*} \phi)(p)=\frac{d}{dt}\phi(\exp(tX)\cdot p)|_{t=0}
\mbox{.}
\eqno{(3.1)}
$$
Since $< , >$\, is $G-$ invariant, one knows that $X^{*}$ is a Killing vector 
field. $X^{*}$ is said to be non-degenerate if for every zero $p\in M$ of
$X^{*}$, the induced linear map ${\frak L}_p(X^{*}): T_p(M)\rightarrow T_p(M)$ 
non-singular. Since $X^{*}$ is a Killing vector field, ${\frak L}_p(X^{*})$
is skew-symmetric with respect to the inner product structure $< , >_p$ on
$T_p(M)$, as we have noted, and non-singularity of ${\frak L}_p(X^{*})$
means that we can construct the square-root

$$
\left[{\rm det}{\frak L}_p(X^{*})\right]^{1/2}=(-1)^n{\rm Pf}_e
({\frak L}_p(X^{*}))=\lambda_1\cdot\cdot\cdot \lambda_n
\mbox{,}
\eqno{(3.2)}
$$
as in (2.14).

For a form $\tau\in \Lambda M =\sum \oplus \Lambda^jM$ we write 
$\tau_j\in \Lambda^jM$ for its homogeneous $j-th$ component,

$$
\tau = (\tau_0,..., \tau_{2n})=\sum_{j=0}^{2n}\tau_j
\mbox{,}
\eqno{(3.3)}
$$
and we write $\left[{\tau}\right]$ for the cohomology class of $\tau$ in case
$\tau\in Z(M,Y,s)$ for $Y\in VM,\, s\in {\Bbb C}$; i.e. $d_{Y,s}\tau=0$
for $d_{Y,s}$ in (2.4). When $M$ is compact, in particular, one can integrate
any $2n-$ form (as $M$ is orientable). Thus we can define

$$
\int_M \tau = \int_M \tau_{2n}
\mbox{,}
\eqno{(3.4)}
$$
and in fact we can define

$$
\int_M [\tau] =\int_M \tau = \int_M \tau_{2n}
\mbox{.}
\eqno{(3.5)}
$$
The integral $\int_M[\tau]$\, really does depend only on the class $[\tau]$\,
of $\tau$. That is, if $\tau'\in B(M,Y,s)$ then by a quick computation
using Stokes' theorem one sees that $\int_M\tau'\stackrel{({\rm i})}{=} 0$.
Similarly if $p\in M$ with $Y_p=0$ then $\tau_0'(p)=0$ for 
$\tau'\in B(M,Y,s)$. In fact if we write $\tau'=d_{Y,s}\beta$ for 
$\beta\in \Lambda_YM$ then one has

$$
\tau'=(si(Y)\beta_1,\, d\beta_0+si(Y)\beta_2,\, d\beta_1+si(Y)\beta_3,\,
d\beta_2+si(Y)\beta_4, 
$$
$$
\!\!\!\!\!\!\!\!\!\!\!\!\!\!\!\!\!\!\!\!\!\!\!\!\!\!\!\!\!\!\!\!
\!\!\!\!\!\!\!\!\!\!\!\!\!\!\!\!\!\!\!\!\!\!\!\!\!\!\!\!\!\!\!\!
...,\, d\beta_{2n-2}+si(Y)\beta_{2n},\, d\beta_{2n-1})
$$
$$
\!\!\!\!\!\!\!\!\!\!\!\!\!
= d\beta_0+si(Y)\beta_0+d\beta_1+si(Y)\beta_1+d\beta_2+si(Y)\beta_2
$$
$$
\!\!\!\!\!\!\!\!\!\!\!\!\!\!\!\!\!\!\!\!\!\!\!\!\!\!\!\!\!\!\!\!\!
\!\!\!\!\!\!\!\!\!\!\!\!\!\!\!\!\!\!\!\!\!\!\!\!\!\!\!\!\!\!\!\!\!
\!\!\!\!\!\!\!\!\!\!\!\!\!
+\, ...\, + d\beta_{2n}+si(Y)\beta_{2n}
\mbox{.}
\eqno{(3.6)}
$$ 
Thus $\tau_0'(p)=s\beta_{1p}(Y_p)=0$, and 
$\int_M\tau'=\int_Md\beta_{2n-1}=0$, which proves $({\rm i})$. It follows that 
the map $p^{*}: H(M,Y,s)\rightarrow {\Bbb R}$ given by

$$
p^{*}[\tau]=\tau_0(p)\,\,\, {\rm for}\,\,\, Y_p=0
\eqno{(3.7)}
$$
is well-defined.

In \cite{berline1,berline2,berline3}, N. Berline and M. Vergne, following 
some ideas of R. Bott in \cite{bott}, established the following localization 
theorem, where the choice $s=-2\pi\sqrt{-1}$ is made.

\medskip
\par \noindent
{\bf Theorem 3.1}.\,\,\,{\em Assume as above that $M$ and $G$ are compact
and that the Riemannian metric $< , >$ on $M$ is $G-$ invariant; i.e.
each $a\in G$ acts as an isometry of $M$. For $X\in {\rm {\bf g}}$, the Lie 
algebra
of $G$, assume that the induced vector field $X^{*}$ on $M$ (see (3.1)) is
non-degenerate; thus the square-root in (3.2) is well-defined (and is non-zero)
for $p\in M$ a zero of $X^{*}$ (i.e. $X_p^{*}=0$). Then for any cohomology
class $[\tau]\in H(M,X^{*},-2\pi\sqrt{-1})$ one has

$$
\int_M[\tau]=(-1)^{n/2}
\!\!\!\!\!
\sum_{\scriptstyle p\in M, 
\atop\scriptstyle p=\,{\rm a\,\, zero\,\, of}\,\,X^{*}}
\frac{p^{*}[\tau]}
{[{\rm det}{\frak L}_p(X^{*})]^{1/2}}
\mbox{;}
\eqno{(3.8)}
$$
see (3.5), (3.7).
}
\medskip

For concrete applications of Theorem 3.1 we shall need to construct concrete 
cohomology classes in $H(M,X^{*}, -2\pi\sqrt{-1})$. The construction of such
classes requires that a bit more be assumed about $M$ and $G$. Suppose for
example that $M$ has a symplectic structure $\sigma: \sigma \in \Lambda^2M$
is a closed 2-form (i.e. $d\sigma=0$) such that for every $p\in M$ the
corresponding skew-symmetric form 
$\omega_p: T_p(M)\oplus T_p(M)\rightarrow {\Bbb R}$ is non-degenerate. In 
particular $M$ is oriented by the Liouville form

$$
\omega_{\sigma}=\frac{1}{n!}\sigma \wedge\cdot\cdot\cdot \wedge\sigma \in
\Lambda^{2n}M - \{0\}
\mbox{.}
\eqno{(3.9)}
$$
Suppose also that there is a map $J: {\rm {\bf g}}\rightarrow C^{\infty}(M)$
which satisfies

$$
i(X^{*})\sigma + dJ(X)=0,\,\,\,\,\,
\forall X\in {\rm {\bf g}}
\mbox{,}
\eqno{(3.10)}
$$
an equality of 1-forms. The existence of such a map $J$ amounts to the 
assumption that the action of $G$ on $M$ is Hamiltonian, a point which we 
shall return to later. Given $J$ define for each $X\in {\rm {\bf g}}$ the form
$\tau^{X}\in \Lambda M$ by

$$
\tau^{X}= \left(J(X), 0, -\frac{\sigma}{2\pi\sqrt{-1}}, 0,..., 0
\right);
\eqno{(3.11)}
$$
see (3.3). We claim that $\tau^{X}\in Z(M,X^{*},-2\pi\sqrt{-1})$. Since
$J(X)$ is a function $i(X^{*})J(X)=0$. Therefore by (2.3) and (3.10),
$\theta(X^{*})J(X)=i(X^{*})dJ(X)=-i(X^{*})^2\sigma = 0$ and 
$\theta(X^{*})\sigma = di(X^{*})\sigma+i(X^{*})d\sigma =di(X^{*})\sigma$
(as $d\sigma=0$) $= -d^2J(X)=0$. By definition (3.11) it follows that
$\theta(X^{*})\tau^X= \left(\theta(X^{*})J(X), 0, 
-\theta(X^{*})\sigma/2\pi\sqrt{-1},0,...,0\right)=0$, which by (2.5) means
that $\tau^X\in \Lambda_{X^{*}}M$. Also for $s=-2\pi\sqrt{-1}$, by
definition (2.4) and (3.10), $d_{X^{*},s}\tau^X = (d+si(X^{*}))\tau^X
= dJ(X)+si(X^{*})J(X)-d\sigma/2\pi\sqrt{-1}-si(X^{*})\sigma/2\pi\sqrt{-1}
= -i(X^{*})\sigma + i(X^{*})\sigma =0$, which verifies the claim, where again
we have used that $i(X^{*})J(X)= 0,\, d\sigma = 0$. Thus, given $J$, we have 
for each $X\in {\rm {\bf g}}$ a cohomology class 
$[\tau^X] \in H(M,X^{*},-2\pi\sqrt{-1})$.

\section{The class $\left[e^{{c\tau}^X}\right]$}

In the next section the Duistermaat-Heckman formula will be derived by a 
direct application of Theorem 3.1. The main point is the construction of an 
appropriate cohomology class. Namely for the cocycle 
$\tau^{X}\in Z(M,X^{*},-2\pi\sqrt{-1})$ in (3.11) we wish to construct for 
$c\in {\Bbb C}$ a well-defined form $e^{{c\tau}^X}$ which also is an element of
$Z(M,X^{*},-2\pi\sqrt{-1})$.

Thus again suppose $J$ which satisfies (3.10) is given. For 
$X\in {\rm {\bf g}}$
let $\tau_0=J(X),\, \tau_1=0,\, \tau_2=-\sigma/2\pi\sqrt{-1},\, \tau_j=0$ for
$3\leq j \leq 2n$, and let $\tau=\tau^X$. That is, by (3.11), 
$\tau =(\tau_0, \tau_1, \tau_2, ..., \tau_{2n})=(\tau_0, 0, 
\tau_2, 0,0,...,0)$. If $\omega_1, \omega_2$ are forms of degree $p,q$
respectively, then $\omega_1$ and $\omega_2$ commute if either $p$ or $q$
is even, since $\omega_1\wedge \omega_2=(-1)^{pq}\omega_2\wedge \omega_1$.
In particular $\tau_0$ and $\tau_2$ commute. Now if $A$ and $B$ are commuting
matrices one has $e^{A+B}=e^A\cdot e^B$. Since $\tau_0$ and $\tau_2$ commute
we should have, formally for any complex number $c,\, c\tau=c\tau_0+c\tau_2
\Rightarrow e^{c\tau}=e^{c\tau_0}\cdot e^{c\tau_2}=e^{c\tau_0}(1+c\tau_2
+c^2\tau_2^2/2!+c^3\tau_2^3/3!+...)$, with $\tau_2^j=\tau_2\wedge\cdot\cdot\cdot\wedge\tau_2$ ($j$ times) $\in \Lambda^{2j}M$. Since $\Lambda^{2j}M=0$ for
$j>n$ we can take $\sum_{j=0}^{\infty}c^j\tau_2^j/j!$ to mean
$\sum_{j=0}^{n}c^j\tau_2^j/j!$. That is, thinking of $c\tau_2^j/j!$ as
$(0,0,...,c\tau_2^j/j!,0,...,0)$ and 1 as $(1,0,0,...,0)$ for 
$1\in C^{\infty}(M)$, we are therefore lead to define $e^{c\tau}$ by

$$
e^{c\tau}=\left(e^{c\tau_0},\,0\,,e^{c\tau_0}c\tau_2,\, 0\,, e^{c\tau_0}
\frac{1}{2!}c^2\tau_2^2,\, 0,\, e^{c\tau_0}\frac{1}{3!}c^3\tau_2^3,\, 0,\, 
\right.
$$
$$
\left.
\!\!\!\!\!\!\!\!\!\!\!\!\!\!\!\!\!\!\!\!\!\!\!\!\!\!\!\!\!
\!\!\!\!\!\!\!\!\!\!\!\!\!\!\!\!\!\!\!\!\!\!\!\!\!\!\!\!\!
...,\,\,0,\,\, e^{c\tau_0}
\frac{1}{n!}c^n\tau_2^n\right) \in \Lambda M
\mbox{;}
\eqno{(4.1)}
$$
compare (3.3). Now $i(X^{*})e^{c\tau_0}=0$ (as $e^{c\tau_0}$ is a function),
and $de^{c\tau_0}=ce^{c\tau_0}d\tau_0$. That is, by (2.3), 
$\theta(X^{*})e^{c\tau_0}=ci(X^{*})e^{c\tau_0}d\tau_0  =
c[i(X^{*})e^{c\tau_0}d\tau_0\,
+e^{c\tau_0}i(X^{*})d\tau_0]= ce^{c\tau_0}i(X^{*})d\tau_0$, where 
$\tau_0=J(X) \Rightarrow$ (by (2.3), (3.10)) $i(X^*)d\tau_0=-i(X^*)^2\sigma =0
\Rightarrow \theta(X^*)e^{c\tau_0}\stackrel{({\rm ii})}{=}0$. More generally,
\,\,\,\,\,\,\,\,\,\,\,\,\,\,\,
$\theta(X^*)e^{c\tau_0}(c^j\tau_2^j)/j!=(\theta(X^*)e^{c\tau_0})
(c^j\tau_2^j)/j! + e^{c\tau_0}(c^j/j!)\theta(X^*)\tau_2^j 
\,\,\,\,\,\,\,\,\,\,\,\,\,\,\,
= e^{c\tau_0}(c^j/j!)\theta(X^*)\tau_2^j$ (by $({\rm ii})$) $= 0$, 
again by the fact 
that $\theta(X^*)$ is a
derivation and the fact that $\theta(X^{*})\tau_2=
-1/2\pi\sqrt{-1}\theta(X^{*})\sigma$ with $\theta(X^{*})\sigma=0$
(as abserved earlier). By (4.1) we see therefore that 
$\theta(X^{*})e^{c\tau}=0 \Rightarrow e^{c\tau}\in \Lambda_{X^{*}}M$, by
(2.5). We claim moreover that $d_{X^*,s}e^{c\tau}=0$ for $s=-2\pi\sqrt{-1}$.
By (3.6) and (4.1)

$$
d_{X^*,s}e^{c\tau}=\left(0,\,d\beta_0+si(X^*)\beta_2,\,0,\,d\beta_2+
si(X^*)\beta_4,\,0, \right.
$$
$$
\left.
\!\!\!\!\!\!\!\!\!\!\!\!\!\!\!\!
\!\!\!\!\!\!\!\!\!\!
...,\,d\beta_{2n-2}+si(X^*)\beta_{2n},\,0\right)
\eqno{(4.2)}
$$
for $\beta_{2j}=e^{c\tau_0}c^j\tau_2^j/j!$. Using that 
$d(\omega_1\wedge\omega_2)\stackrel{({\rm iii})}{=}d\omega_1\wedge\omega_2+
(-1)^{{\rm deg}\omega_1}\omega_1\wedge d\omega_2$ for forms $\omega_1,\omega_2$
of homogeneous degree and that $e^{c\tau_0}, \tau_2$ are of even degree, we get
$de^{c\tau_0}\tau_2^j=de^{c\tau_0}\wedge\tau_2^j + e^{c\tau_0}\wedge d\tau_2^j$
where $d\tau_2^j=0$ (by $({\rm iii})$ since 
$d\tau_2=-1/(2\pi\sqrt{-1})d\sigma=0)$
$\Rightarrow d\beta_{2j}=(c^j/j!)e^{c\tau_0}dc\tau_0\wedge\tau_2^j
\stackrel{({\rm iv})}{=}-(c^{j+1}/j!)e^{c\tau_0}(i(X^*)\sigma)\wedge\tau_2^j$,
by (3.10). Similarly $i(X^*)e^{c\tau_0}\tau_2^j=(i(X^*)e^{c\tau_0})\tau_2^j
+ e^{c\tau_0}i(X^*)\tau_2^j= e^{c\tau_0}i(X^*)\tau_2^j$, where
$i(X^*)\tau_2^j = j\tau_2^{j-1}\wedge i(X^*)\tau_2$ (since $i(X^*)$ also 
satisfies the derivative property $({\rm iii})$, and since $i(X^*)\tau_2$ and
$\tau_2$
commute as ${\rm deg}\,\tau_2=2$) $\Rightarrow i(X^*)\beta_{2j}
e^{c\tau_0}(c^j/(j-1)!)\tau_2^{j-1}\wedge i(X^*)\tau_2\,\, = 
e^{c\tau_0}(c^j/(j-1)!)\tau_2^{j-1}\wedge i(X^*)\sigma/s$\,
(for $s=-2\pi\sqrt{-1}$) $\Rightarrow si(X^*)\beta_{2j+2}$
$\stackrel{{\rm v}}{=}\,
e^{c\tau_0}(c^{j+1}/j!)\tau_2^j\wedge i(X^*)\sigma$. That is, by $({\rm iv})$
and ({\rm v}), $d\beta_{2j}+si(X^*)\beta_{2j+2}=0$ (again as $i(X^*)\tau_2$
and $\tau_2$ commute), which by (4.2) establishes the claim. Hence the 
following is proved.

\medskip
\par \noindent

{\bf Theorem 4.1}.\,\,\,{\em Suppose $J: {\rm {\bf g}}\rightarrow 
C^{\infty}(M)$
which satisfies (3.10) is given, where $\sigma$ is a symplectic structure on
$M$. Recall that for $X\in {\rm {\bf g}}$, equation (3.11) defines a cocycle 
$\tau^X\in Z(M,X^*, -2\pi\sqrt{-1})$. Similarly for $c\in {\Bbb C}$,
define $e^{c\tau^X}$ by (4.1):

$$
e^{c\tau^X}=\left(e^{cJ(X)}, 0, e^{cJ(X)}c\left(\frac{\sigma}{-2\pi\sqrt{-1}}
\right), 0, e^{cJ(X)}\frac{c^2}{2!}\left(\frac{\sigma}{-2\pi\sqrt{-1}}
\right)^2, 0, 
\right.
$$
$$
\left.
\!\!\!\!\!\!\!\!\!\!\!\!\!\!\!\!\!\!\!\!\!\!\!\!
\!\!\!\!\!\!\!\!\!\!\!\!\!\!\!\!\!\!\!\!\!\!\!\!
...,\,\, 0,\,\, e^{cJ(X)}\frac{c^n}{n!}\left(\frac{\sigma}{-2\pi\sqrt{-1}}
\right)^n\right)\in \Lambda M
\mbox{,}
\eqno{(4.3)}
$$
for ${\rm dim}\,M=2n$. Then also $e^{c\tau^X}\in Z(M,X^*, -2\pi\sqrt{-1})$,
and thus we have the cohomology class $\left[e^{c\tau^X}\right] \in 
H(M, X^*, -2\pi\sqrt{-1})$; see (2.4), (2.6), (3.1).}

\section{The Duistermaat-Heckman Formula}

Theorem 4.1 contains the basic assumption that a function 
$J: {\rm {\bf g}}\rightarrow C^{\infty}(M)$ exists which satisfies 
condition (3.10).
As pointed out earlier this assumption amounts to the assumption that the
action of $G$ on $M$ is Hamiltonian - a point which we will now explain.

Given the symplectic structure $\sigma$ on $M$ there is a duality 
$Y\leftrightarrow \beta_Y$ between smooth vector fields $Y\in VM$ and smooth
1-forms $\beta_Y\,\,\, \Lambda^1M$ on $M$:

$$
\beta_Y(Z)=\sigma(Y,Z)\,\,\,\,\,\,\,{\rm for\,\,\,every}\,\,\, Z\in VM
\mbox{.}
\eqno{(5.1)}
$$
$Y\in VM$ is called a Hamiltonian vector field if $\beta_Y$ is exact:
$\beta_Y = d\phi$ for some $\phi\in C^{\infty}(M)$. Let $HVM$ denote the
space of Hamiltonian vector fields on M. Actually $HVM$ is a Lie algebra.
For example, given any $\phi\in C^{\infty}(M)$, the smooth 1-form $d\phi$
corresponds (by the aforementioned duality) to a smooth vector field $Y_\phi$
on $M$. Thus $Y_\phi\in HVM$ and by (2.2) and (5.1) we have for every
$Z\in VM, \, (i(Y_\phi)\sigma)(Z)=\sigma(Y_\phi,Z)=d\phi(Z)\, \Rightarrow$

$$
d\phi=i(Y_\phi)\sigma
\mbox{.}
\eqno{(5.2)}
$$
The equation

$$
[\phi_1, \phi_2]= Y_{\phi_1}\phi_2\,\,\,\,\,{\rm for}\,\,\,
\phi_1,\phi_2\in C^{\infty}(M)
\eqno{(5.3)}
$$
defines the Poisson bracket $[\,,\,]$ on $C^{\infty}(M)$ which converts 
$C^{\infty}(M)$ into a Lie algebra such that the map\,\, 
$\wp : \phi\rightarrow Y_\phi: C^{\infty}(M)\rightarrow HVM$ is a Lie
algebra homomorphism; i.e. $[Y_{\phi_1},Y_{\phi_2}]=Y_{[\phi_1,\phi_2]}$.
The (left) action of $G$ on $M$ is called symplectic if $X^*\in HVM,\,\,\, 
\forall X\in {\rm {\bf g}}$; see (3.1). Now the map $X\rightarrow X^*: 
\,\,{\rm {\bf g}}\rightarrow VM$ is not a Lie algebra homomorphism since
$[X_1,X_2]^*=-[X_1^*, X_2^*]$ for $X_1, X_2 \in {\rm {\bf g}}$. If we define 
$\eta: {\rm {\bf g}}\rightarrow VM$ by $\eta(X)=(-X^*)=-X^*$ then we do 
obtain a 
homomorphism: $\eta([X_1, X_2])=-[X_1, X_2]^* = [X_1^*, X_2^*]=
[-\eta(X_1), -\eta(X_2)] = [\eta(X_1), \eta(X_2)]$. In other words if the
action of $G$ is symplectic then $\eta: {\rm {\bf g}}\rightarrow HVM$ is a Lie 
algebra homomorphism. The (left) action of $G$ on $M$ is called Hamiltonian
if it is symplectic and if the Lie algebra homomorphism 
$\eta: {\rm {\bf g}}\rightarrow HVM$ has a lift to $C^{\infty}(M)$ - i.e. if
there exists a Lie algebra homomorphism $J: {\rm {\bf g}}\rightarrow C^{\infty}(M)$
such that the diagram
\vspace{1.0cm}
\medskip
\par \noindent
$$
\begin{picture}(120,80)
\put(60,25){${\bf {\rm {\bf g}}}$}
\put(0,80){${\rm C}^{\infty}({\rm M})$}
\put(100,80){${\rm HVM}$}
\put(60,90){$\wp$}
\put(40,83){\vector(1,0){52}}
\put(15,45){{\rm J}}
\put(50,40){\vector(-1,1){30}}
\put(75,40){\vector(1,1){30}}
\put(100,45){$\eta$}
\put(210,50){$(5.4)$}
\end{picture}
$$
is commutative: $\eta=\wp\circ J$, or

$$
-X^* =Y_{J(X)}\,\,\,\,\,{\rm for\,\,\,every}\,\,\,X\in {\bf {\rm {\bf g}}}
\mbox{.}
\eqno{(5.5)}
$$
We note that such a $J$ will indeed satisfy condition (3.10). Namely, by (5.2)
and (5.5), $dJ(X)=i(Y_ {J(X)})\sigma=-i(X^*)\sigma$ for 
$X\in {\bf {\rm {\bf g}}}$. The
triple $(M,\sigma,J)$, for $J$ subject to (5.4), is called a Hamiltonian
$G-$ space \cite{kostant,woodhouse}. The basic example of a Hamiltonian 
$G-$ space is that
of an orbit ${\mathcal O}$ in the dual space ${\bf {\rm {\bf g}}}^*$ of 
${\bf {\rm {\bf g}}}$ under the
adjoint action of $G$ on ${\bf {\rm {\bf g}}}^*$, where $\sigma$ is chosen 
as the Kirillov
symplectic form on $M={\mathcal O}$, and where $J$ is given by a canonical
construction (see Appendix).

We are now in position to state the Duistermaat-Heckman formula - in a form 
directly derivable from Theorem 3.1.

\medskip
\par \noindent

{\bf Theorem 5.1}.\,\,\,{\em Suppose as above that $(M,\sigma,J)$ is a
Hamiltonian $G-$ space where $G$ and $M$ are compact. Orient $M$ by the
Liouville form $\omega_\sigma$ in (3.9). Then for $c\in {\Bbb C}$ and for
$X\in {\bf {\rm {\bf g}}}$ with $X^*$ non-degenerate, we have

$$
\int_Me^{cJ(X)}\omega_\sigma = \left(\frac{2\pi}{c}\right)^n
\sum_{\scriptstyle p\in M, 
\atop\scriptstyle p=\,{\rm a\,\, zero\,\, of}\,\,X^{*}}
\frac{e^{cJ(X)(p)}}{\left[{\rm det}\,{\frak L}_p(X^*)\right]^{\frac{1}{2}}}
\mbox{.}
\eqno{(5.6)}
$$
} Here, as in Theorem 3.1, some $G-$ invariant Riemannian metric $<,>$ on
$M$ has been selected, and the square-root in (5.6) is that in (3.2).

The proof of (5.6) is quite simple, given Theorem 3.1. Namely, given the 
lifting $J$ (where we have noted that (5.4) implies (3.10)) let
$c_J(X)=\left[e^{c\tau^X}\right]$ be the cohomology class constructed in
Theorem 4.1, for $c\in {\Bbb C},\,\, X\in {\bf {\rm {\bf g}}}$. By (3.7) 
and (4.3)

$$
p^*c_J(X)=e^{cJ(X)(p)}\,\,\,\,\,{\rm for}\,\,\,X_p^*=0
\mbox{,}
\eqno{(5.7)}
$$
and by (3.5) and (4.3)

$$
\int_Mc_J(X) = \left(\frac{c}{-2\pi\sqrt{-1}}\right)^n\int_Me^{cJ(X)}
\frac{\sigma^n}{n!}
= (-1)^{\frac{n}{2}}\left(\frac{c}{2\pi}\right)^n\int_Me^{cJ(X)}
\omega_\sigma
\mbox{.}
\eqno{(5.8)}
$$
On the other hand given that $X^*$ is non-degenerate, the localization
formula (3.8) gives

$$
\int_Mc_J(X) = (-1)^{\frac{n}{2}}
\sum_{\scriptstyle p\in M, 
\atop\scriptstyle p=\,{\rm a\,\, zero\,\, of}\,\,X^{*}}
\frac{e^{cJ(X)(p)}}{\left[{\rm det}\,{\frak L}_p(X^*)\right]^{\frac{1}{2}}}
\mbox{,}
\eqno{(5.9)}
$$
by (5.7). That is, by (5.8) and (5.9) we obtain exactly formula (5.6), as 
desired.

Note that for $X\in {\rm {\bf g}},\, Z\in VM$, and 
$p\in M,\,\,\, \, dJ(X)_p(Z_p) = [dJ(X)(Z)](p)\,\,\, =\,\,\, 
[(-i(X^*)\sigma)(Z)](p)$
\,\,\,(as $J$ satisfies (3.10))\,\,\, $=$\,\,\,\,\,\, 
$-\sigma(X^*, Z)(p)$\,\, (by (2.2))\,\,\, $= -\sigma_p(X_p^*, Z_p)$.
Hence $dJ(X)_p=0$ if $X_p^* = 0$, and conversely $dJ(X)_p = 0$\,\,\,
$\Rightarrow\,\,\,X_p^* = 0$ since $\sigma_p$ is non-degenerate. (5.6) can
therefore be expressed as

$$
\int_Me^{cJ(X)}\omega_\sigma = \left(\frac{2\pi}{c}\right)^n
\sum_{\scriptstyle p\in M, 
\atop\scriptstyle p=\,{\rm a\,\, critical\,\,point\,\, of}\,\,J(X)}
\frac{e^{cJ(X)(p)}}{\left[{\rm det}\,{\frak L}_p(X^*)\right]^{\frac{1}{2}}}
\mbox{,}
\eqno{(5.10)}
$$
where the critical points of $J(X)$ are those where $dJ(X)$ vanishes.
Recall that the asymptotic behaviour of an oscillatory integral

$$
I(f, t) = \int_{X (={\rm some\,\,space})}e^{\sqrt{-1}tf(x)}dx
\eqno{(5.11)}
$$
for large $t$ is given by the stationary-phase approximation - the dominant
terms of this approximation being governed by the critical points of the phase
$f(x)$. If we choose $c=\sqrt{-1}t$, for $t\in {\Bbb R}$, in (5.10), in
particular, we see that the D-H formula can be viewed as an 
exactness result in a stationary-phase approximation of the integrals
$\int_Me^ {\sqrt{-1}tJ(X)}\omega_\sigma$, as our remarks of Section 1 
indicated.

For extended and much broader discussions of material introduced here, the two
references \cite{berline4,szabo} are especially recommended. The reference 
\cite{szabo} in particular serves as a vast source of information for the 
needs of physicists. Further reading of interest is found in the references
\cite{atiyah2,cordes,dykstra,funahashi,niemi2,niemi3,niemi4,paradan,picken,
schwarz,stone,witten1,witten2,witten3}.

\section{Appendix}

The D-H formula of Theorem 5.1 was stated in the context of a Hamiltonian $G-$
space $(M,\sigma,J)$. We pointed out that the premier example of such a space
is an orbit ${\mathcal O}$ in the dual space ${\rm {\bf g}^*}$ of the Lie
algebra ${\rm {\bf g}}$ of a Lie group $G$, where the action of $G$ on
${\rm {\bf g}^*}$ (which is called the co-adjoint action) is induced by the 
adjoint action of $G$ on ${\rm {\bf g}}$. Namely for a linear functional
$f$ on ${\rm {\bf g}}$,\, $f\in {\rm {\bf g}^*}$,

$$
(a\cdot f)(X) = f(Ad(a^{-1})X)\,\,\,\,\,{\rm for}\,\,\,a\in G,\,\,
X\in {\rm {\bf g}}
\mbox{.}
\eqno{(A.1)}
$$
We shall recall how the (well-known) symplectic structure $\sigma$ on
${\mathcal O}$ is obtained (due to A.A. Kirillov) and how the lifting $J$ is
canonically constructed. Thus we exhibit $({\mathcal O}, \sigma=
\sigma_{\mathcal O}, J=J_{\mathcal O})$ as a key example of a 
Hamiltonian $G-$ space. For this purpose 
it is convenient to regard the orbit of $f$ as a homogeneous space:
${\mathcal O} \simeq G/G_f$ where $G_f$ is the stabilizer of $f$:

$$
G_f = \{ a\in G | a\cdot f = f \}
\mbox{.}
\eqno{(A.2)}
$$
$G_f$ is a closed subgroup of $G$ with Lie algebra ${\rm {\bf g}_f}$
given by

$$
{\rm {\bf g}_f} = \{X \in {\rm {\bf g}}\,\, |\, f([X, Y]) =0
\,\,\,\,\, \forall\, Y\in {\rm {\bf g}} \}
\mbox{.}
\eqno{(A.3)}
$$

Let $\tau^f$ be the corresponding Maurer - Cartan 1-form on $G$. That is,
$\tau^f \in V^1G$ is the unique left-invariant 1-form on $G$ subject to the
condition

$$
\tau^f(X)(1) = f(X)\,\,\,\,\,\,\,\,\,\,\, \forall \,X \in {\rm {\bf g}}
\mbox{.}
\eqno{(A.4)}
$$
Let $\pi: \,G\rightarrow G/G_f$ denote the quotient map.

\medskip
\par \noindent
{\bf Theorem A.1}.\,\,\,{\em $G/G_f$ has a symplectic structure $\sigma$
which is uniquely given by $\pi^*\sigma = d\tau^f$.}
\medskip
\medskip

Here $\pi^*\omega_1$ denotes the pull-back of a form $\omega_1$. The form
$\sigma$ is also left-invariant; i.e. $\ell_a^*\sigma = \sigma$ where
$\ell_a: G/G_f \rightarrow  G/G_f$ denotes left translation by $a\in G$.
Given $X \in {\rm {\bf g}}$ define $\phi_X: G/G_f \rightarrow {\Bbb R}$ by

$$
\phi_X(aG_f) = f(Ad(a^{-1})X) = (a\cdot f)(X)
\eqno{(A.5)}
$$
for $a \in G$; $\phi_X$ is well-defined by (A.2). One can show by computation
that

$$
d\phi_X = -i(X^*)\sigma
\mbox{.}
\eqno{(A.6)}
$$
That is, by (5.1), $\beta_{-X^*} = d\phi_X \Rightarrow -X^*$ (or $X^*$) is
Hamiltonian for each $X \in {\rm {\bf g}}$; i.e. the action of $G$ on
$G/G_f$ is symplectic. To see that this action is Hamiltonian we must construct
a lift $J: {\rm {\bf g}}\Rightarrow C^{\infty}(G/G_f)$ of 
$\eta: X\rightarrow -X^*$. Namely define $J$ by

$$
J(X) = \phi_X \,\,\,\,\,\, {\rm for}\,\,\,\,\phi_X\,\,\, {\rm in}\,\,\, 
(A.5)
\mbox{.}
\eqno{(A.7)}
$$
Recall that $\wp: C^{\infty}(M) \rightarrow HVM$ is given by 
$\wp(\phi) = Y_\phi$. That is, by (5.2) and (A.6), 
$\wp(\phi_X) = - X^* = \eta (X)$, which shows that $J$ does satisfy the 
commutative diagram in (5.4). The final step is to show that $J$ is a 
homomorphism. Let $X_1, X_2 \in {\rm {\bf g}}$, $a \in G$. The Poisson bracket
is given by (5.3):

$$
\left[J(X_1), J(X_2)\right](\pi(a)) = \left(Y_{J(X_1)}J(X_2)\right)(\pi(a))
$$
$$
= \left(\wp(J(X_1)) J(X_2)\right)(\pi(a)) = \left(\eta(X_1)J(X_2)\right)
(\pi(a))
$$
$$
({\rm again\,\,\, by}\,\,\, (5.4)) = \left((-X_1^*)\phi_{X_2}\right)(\pi(a))
\,\,\,({\rm by}\,\,\,(A.7))
$$
$$
= \frac{d}{dt}\phi_{X_2}\left((\exp (-tX_1))\cdot \pi(a)\right)|_{t=0}
\,\,\,\,\,({\rm by}\,\,\,\, (3.1))
$$
$$
= \frac{d}{dt}\phi_{X_2}\left(\pi((\exp (-tX_1))\cdot a)\right)|_{t=0}
$$
$$
= \frac{d}{dt}f\left(Ad(a^{-1}\exp(X_1))X_2\right)|_{t=0}
\,\,\,\,\,({\rm by}\,\,\,\, (A.5))
$$
$$
= \frac{d}{dt}f\left(Ad(a^{-1})Ad(\exp(X_1))X_2\right)|_{t=0}
$$
$$
= \frac{d}{dt}(a\cdot f)\left(Ad(\exp(X_1))X_2\right)|_{t=0}
\,\,\,\,\,({\rm by}\,\,\,\, (A.5))
$$
$$
= (a\cdot f)\left([X_1, X_2]\right) = f\left(Ad(a^{-1})[X_1, X_2]\right)
\mbox{.}
\eqno{(A.8)}
$$
On the other hand

$$
J\left([X_1, X_2]\right)(\pi(a)) = \phi_{[X_1, X_2]}(\pi(a)) 
\,\,\,\,\, ({\rm by}\,\,\,(A.7))
$$
$$
= f\left(Ad(a^{-1})[X_1, X_2]\right)\,\,\,\,\, ({\rm by}\,\,\, (A.5))
\eqno{(A.9)}
$$
which proves that $[J(X_1), J(X_2)] = J([X_1, X_2])$.


\begin{thebibliography}{10}

\bibitem{atiyah1}
{\sc M. Atiyah}, {\em ``Circular Symmetry and Stationary - Phase 
Approximation''}, Colloquim in Honor of Laurent Schwartz 1, Ast\'{e}risque
{\bf 131} (1985) 43-59.

\bibitem{atiyah2}
{\sc M. Atiyah and R. Bott}, {\em ``The Moment Map and Equivariant 
Cohomology''}, Topology {\bf 23}(1) (1984) 1-28.

\bibitem{berline1}
{\sc N. Berline and M. Vergne}, {\em ``Classes Charact\'{e}ristiques 
\'{E}quivariantes, Formules de Localisation en Cohomologie \'{E}quivariante''},
C.R. Acad. Sci. Paris {\bf 295} (1982) 539-541.

\bibitem{berline2}
{\sc N. Berline and M. Vergne}, {\em ``Z\'{e}ros d'un Champ de Vectors
et Classes Charact\'{e}ristiques \'{E}quivariantes''}, Duke Math. J.
{\bf 50} (1983) 539-549.

\bibitem{berline3}
{\sc N. Berline and M. Vergne}, {\em ``Fourier Transforms of Orbits of 
the Co-adjoint Representation''}, In Proceedings of the Conference on 
Representation Theory of Reductive Groups (Park City, Utah, 1982), Progress
in Math., Birkh\"{a}user, Boston {\bf 40} (1983) 53-67.

\bibitem{berline4}
{\sc N. Berline, E. Getzler and M. Vergne}, {\em ``Heat Kernels and Dirac
Operators''}, Springer - Verlag, Berlin (1991).

\bibitem{bismut1}
{\sc J. Bismut}, {\em ``Index Theorem and Equivariant Cohomology on the Loop
Space''}, Commun. Math. Phys. {\bf 98} (1985) 213-237.

\bibitem{bismut2}
{\sc J. Bismut}, {\em ``Localization Formulas, Superconnections, and the Index
Theorem for Families''}, Commun. Math. Phys. {\bf 103} (1986) 127-166.

\bibitem{blau}
{\sc M. Blau and G. Thompson}, {\em ``Localization and Diagonalization: A 
Review of Functional Integral Techniques for Low - Dimensional Gauge
Theories and Topological Field Theories''}, J. Math. Phys. {\bf 36} (5) 
(1995) 2192-2236.

\bibitem{bott}
{\sc R. Bott}, {\em ``Vector Fields and Characteristic Numbers''}, 
Mich. Math. J. {\bf 14} (1967) 231-244.

\bibitem{cordes}
{\sc S. Cordes, G. Moore and S. Ramgoolam}, {\em ``Lectures on 2D Yang-Mills 
Theory, Equivariant Cohomology and Topological Field Theories''}, 
In Fluctuating Geometries in Statistical Mechanics and Field Theory (Les 
Houches), North Holland, Amsterdam (1996) 505-682.

\bibitem{duistermaat}
{\sc J. Duistermaat and G. Heckman}, {\em ``On the Variation in the Cohomology 
of the Symplectic Form of the Reduced Phase Space''}, 
Invent. Math. {\bf 69} (1982) 259-268.

\bibitem{dykstra}
{\sc H. Dykstra, J. Lykken and E. Raiten}, {\em ``Exact Path Integrals by 
Equivariant Localization''}, Phys. Lett. {\bf B 302} (1993) 223-229.
 
\bibitem{funahashi}
{\sc K. Funahashi, T. Kashiwa, S. Sakoda and K Fujii}, 
{\em ``Exactness in the WKB Approximation for some Homogeneous Spaces''}, 
J. Math. Phys. {\bf 36} (1995) 4590-4611.
 
\bibitem{kostant}
{\sc B. Kostant}, {\em ``Quantization and Unitary Representations''}, 
In Lectures in Modern Analysis and Applications III, Lecture Notes in
Math., Springer-Verlag {\bf 170} (1970) 87-208.

\bibitem{niemi1}
{\sc A. Niemi}, {\em ``Localization, Equivariant Cohomology and Integration 
Formulas''}, In Particles and Fields, CRM Series in Math. Phys., Springer 
(1999) 211-250.

\bibitem{niemi2}
{\sc A. Niemi and P. Pasanen}, {\em ``Orbit Geometry, Group Representations,
and Topological Quantum Field Theories''}, Phys. Lett. {\bf B 253} (1991)
349-356.

\bibitem{niemi3}
{\sc A. Niemi and O. Tirkkonen}, {\em ``Cohomological Partition Functions
for a Class of Bosonic Theories''}, Phys. Lett. {\bf B 293} (1992)
339-343.

\bibitem{niemi4}
{\sc A. Niemi and O. Tirkkonen}, {\em ``On Exact Evaluation of Path 
Integrals''}, Ann. of Phys. {\bf B 235} (1994) 318-349.

\bibitem{paradan}
{\sc P. Paradan}, {\em ``Action Hamiltoniene d'un Tore et Formula de
Localisation en Cohomologies Equivariante''}, C.R. Acad. Sci. Paris 
{\bf 324} (1997) 491-496.

\bibitem{picken}
{\sc R. Picken}, {\em ``The Duistermaat-Heckman Integration Formula on
Flag Manifolds''}, J. Math. Phys. {\bf 31} (1990) 616-638.

\bibitem{schwarz}
{\sc A. Schwarz and O. Zaboronsky}, {\em ``Supersymmetry and Localization''}, 
Commun. Math. Phys. {\bf 183} (1997) 463-476.

\bibitem{semenov}
{\sc M. Semenov-Tjan-Shanskii}, {\em ``A Certain Property of the Kirillov
Integral''}, In Differential Geometry, Lie Groups, and Mechanics, Math. Ind.
Steklov (LOMI) {\bf 37} (1973) 53-65. 

\bibitem{stone}
{\sc M. Stone}, {\em ``Supersymmetry and the Quantum Mechanics of Spin''}, 
Nucl. Phys. {\bf B 314} (1989) 557-586. 

 \bibitem{szabo}
{\sc R. Szabo}, {\em ``Equivariant Cohomology and Localization of Path
Integrals''}, Lecture Notes in Phys., Springer, m.{\bf 63} (2000). 

\bibitem{witten1}
{\sc E. Witten}, {\em ``Topological Quantum Field Theory''}, 
Commun. Math. Phys. {\bf 117} (1988) 353-386. 

\bibitem{witten2}
{\sc E. Witten}, {\em ``Introduction to Cohomological Field Theory''}, 
Inter. J. Mod. Phys. {\bf A 6} (1991) 2775-2792. 

\bibitem{witten3}
{\sc E. Witten}, {\em ``Two Dimensional Gauge Theories Revisted''}, 
J. Geom. Phys. {\bf 9} (1992) 303-368. 

\bibitem{woodhouse}
{\sc N. Woodhouse}, {\em ``Geometric Quantization''}, 
Clarendon Press (Oxford) (1980). 



\end{thebibliography}
\end{document}